\documentclass[namedreferences,optionalrh]{spr-sola} 
\usepackage{graphicx} 
\usepackage{epsfig} 
\usepackage{enumitem}
\usepackage{siunitx}
\usepackage{multirow}
\usepackage{acronym}

\usepackage{amssymb}                    
\usepackage{xcolor}
\usepackage[backref=page]{hyperref}
\hypersetup{
    colorlinks = true,
    linkcolor = blue,
    anchorcolor = blue,
    citecolor = blue,
    filecolor = blue,
    urlcolor = blue
    }
\graphicspath{{figures/}}

\definecolor{greentheo}{rgb}{0.15,0.50,0.30}
\newcommand{\tobs}{${\rm T}_{\rm OBS}$}

\newcommand{\bkj}[1]{\textcolor{blue}{{\bf #1}}}

\newcommand{\ca}{\mbox{Ca\,{\sc ii}~K\,}}

\begin{document}
\acrodef{utc}[UTC]{Coordinated Universal Time}
\acrodef{ist}[IST]{Indian Standard Time}
\acrodef{dst}[DST]{Daylight Saving Time}
\acrodef{fov}[FOV]{field of view}

\begin{frontmatter}

\title{Butterfly Diagram and other properties of plage areas from Kodaikanal \ca\ Photographs Covering 1904\,--\,2007 }

%
 \author[addressref={aff1,aff2,aff3}, corref,email={maitraibibhu@gmail.com}]{\inits{B.~K.}\fnm{Bibhuti~Kumar}~\lnm{Jha}\orcid{0000-0003-3191-4625}}
 
 \author[addressref={aff4}, corref,email={chatzistergos@mps.mpg.de}]{\inits{T.}\fnm{Theodosios}~\lnm{Chatzistergos}\orcid{0000-0002-0335-9831}}
 
 \author[addressref={aff2,aff3,aff5}]
 {\inits{D.}\fnm{Dipankar}~\lnm{Banerjee}\orcid{0000-0003-4653-6823}}
 
 \author[addressref={aff6}]{\inits{I.}\fnm{Ilaria }~\lnm{Ermolli}\orcid{0000-0003-2596-9523}}
 
 \author[addressref={aff4}]{\inits{N. A.}\fnm{Natalie~A. }\lnm{Krivova}\orcid{0000-0002-1377-3067}}
 
 \author[addressref={aff4}]{\inits{S. K.}\fnm{Sami K. }\lnm{Solanki}\orcid{0000-0002-3418-8449}}
 
 \author[addressref={aff2,aff3}]{\inits{A.}\fnm{Aditya }\lnm{Priyadarshi}\orcid{0000-0003-2476-1536}}

%
\runningauthor{B~K~Jha et al.}
\runningtitle{Updated \ca butterfly diagram from KoSO}

\address[id=aff1]{Southwest Research Institute, Boulder, CO 80302, USA}
\address[id=aff2]{Indian Institute of Astrophysics, Koramangala, Bangalore 560034, India}
\address[id=aff3]{Aryabhatta Research Institute of Observational Sciences, Nainital 263000, India}
\address[id=aff4]{Max Planck Institute for Solar System Research, Justus-von-Liebig-Weg 3, D-37077 Göttingen,Germany}
\address[id=aff5]{Center of Excellence in Space Sciences India, IISER Kolkata, Mohanpur 741246, West Bengal, India}
\address[id=aff6]{INAF Osservatorio Astronomico di Roma, Via Frascati 33, 00078 Monte Porzio Catone, Italy}

\begin{abstract}
\sloppy

\ca observations of the Sun have a great potential for probing the Sun's magnetism and activity, as well as for reconstructing solar irradiance. The Kodaikanal Solar Observatory (KoSO) in India, houses one of the most prominent \ca archives, spanning from 1904 to 2007, obtained under the same experimental conditions over a century, a feat very few other sites have achieved. However, the KoSO \ca\ archive suffers from several inconsistencies (e.g., missing/incorrect timestamps of observations and orientation of some images) which have limited the use of the archive. 
This study is a step towards bringing the KoSO archive to its full potential.
We did this by developing an automatic method to orient the images more accurately than in previous studies. 
Furthermore, we included more data than in earlier studies (considering images that could not previously be analyzed by other techniques as well as 2845 newly digitized images), while also accounting for mistakes in the observational date/time. 
These images were accurately processed to identify plage regions along with their locations, enabling us to construct the butterfly diagram of plage areas from the entire KoSO \ca\ archive covering 1904-–2007.
Our butterfly diagram shows significantly fewer data gaps compared to earlier versions due to the larger set of data used in this study. 
Moreover, our butterfly diagram is consistent with Spörer's law for sunspots, validating our automatic image orientation method. 
Additionally, we found that the mean latitude of plage areas calculated over the entire period is $20.5\%\pm2.0$ higher than that of sunspots, irrespective of the phase or the strength of the solar cycle. We also studied the North-South asymmetry showing that the northern hemisphere dominated plage areas during solar cycles 19 and 20, while the southern hemisphere dominated during solar cycles 21--23.
\end{abstract}

\keywords{Chromosphere, Quiet; Chromosphere, Active;  Solar Cycle, Observations; }
\end{frontmatter}

%
\section{Introduction}
\label{s:intro}
\sloppy

The discovery of two dominant absorption lines of singly ionized Calcium (Ca\,{\sc ii}) in the solar spectrum by Joseph von Fraunhofer in 1814 opened a new window to multi-wavelength solar observation. These two lines, known as H and K lines, probe the solar atmosphere from the photosphere to the chromosphere and serve as excellent diagnostics of the chromospheric structure and dynamics \citep[e.g.][]{Linsky1970,murabito_investigating_2023} as well as of the solar and stellar chromospheric activity \citep[e.g.][]{eberhard_reversal_1913,wilson_flux_1968,sowmya_modeling_2021}. Moreover, solar full disk \ca observations provide unique information on chromospheric plage and network regions, which are observed overlying photospheric magnetic features, such as faculae and network magnetic elements, and hence act as keys for studying the solar surface magnetic field's evolution \citep[e.g.][]{Skumanich1975, Schrijver1989, Harvey1999, Loukitcheva2009, Kahil2017,Chatzistergos2019c}, in particular when magnetograms are not available. Therefore, \ca\ observations are an important asset for studying the evolution of plage \citep[e.g.][]{Chatterjee2016,ermolli_potential_2018,chatzistergos_historical_2019,Chatzistergos2020, chatzistergos_scrutinising_2022,chatzistergos_is_2022,Singh2022}, network \citep{Ermolli1998,ermolli_rome_2022, Berrilli1999, Chatterjee2017},  and reconstructing solar magnetism \citep[e.g.][and references therein]{Chatzistergos2019c,Chatzistergos2022, Mordvinov2020}. Furthermore, since the evolution of the solar surface magnetic field affects the Sun's radiative output, \ca\ observations are also very useful for reconstruction of past total solar irradiance variations \citep[e.g.][]{chatzistergosmodelling2020,Chatzistergos2021,chatzistergos_understanding_2023}.

Although the existence of absorption lines in the solar spectrum was known from the beginning of the 19th century \citep[e.g.][]{secchi_sulla_1871,tacchini_macchie_1875,ermolli_legacy_2021,ermolli_solar_2023},  systematic and photographic observations of the Sun over specific wavelength intervals started only after the invention of spectroheliograph \citep{Hale1890, Hale1891, Hale1893, Deslandres1908}. Since then, synoptic observations of the Sun in \ca have been performed from many sites around the globe \citep[see][and references therein]{Chatzistergos2020b, Chatzistergos2022}, such as those at Kodaikanal \citep[since 1904; ][]{Priyal2014}, Mt. Wilson \citep[since 1915; ][]{lefebvre_solar_2005}, Mitaka \citep[since 1917; ][]{Hanaoka2013}, Arcetri \citep[since 1931; ][]{Ermolli2009} and Meudon \citep[since 1893; ][]{Malherbe2022,malherbe_130_2023}. The fact that  these observations were until recently available only in the form of physical photographs, has been a hindrance to analysing them. This changed dramatically over the last decades, during which most of the existing \ca archives were digitized. By now there is a wealth of digitally available solar \ca data \citep[for a review see][]{Chatzistergos2022}. Among the available archives, Kodaikanal Solar Observatory \citep[KoSO;][]{Hasan2010} has one of the longest records of solar observations in \ca\ covering more than 100~years (1904\,--\,2007). The fact that essentially the same instrumental setup was used for the entire period of observation, makes this dataset an important resource for investigating solar processes over a century-long time scale.

Indeed, the KoSO \ca data have been used for various studies, with determining plage areas being the most common one \citep[e.g.][]{Chatterjee2016,Chatzistergos2019b,Singh2022}. The majority of the past analyses determined disk-integrated quantities and thus did not require any specific orientation of the solar disk. However, there are several applications for which it is essential to know quite accurately the orientation of the solar disk in the images. This includes studies of e.g., the chromospheric solar differential rotation \citep{Singh1985,mishra_differential_2023,kharayat_equator_2024}, the polar network index \citep{Priyal2014b,mishra_ca_2024}, as well as deriving time-latitude maps, commonly referred to as ``butterfly" diagrams useful for studying the evolution of the solar activity and the properties of magnetic regions.

To our knowledge only a few such diagrams have been produced from \ca data \citep{harvey_cyclic_1992,Ermolli2009, Chatterjee2016,Priyal2017,Bertello2020}, two of which were with KoSO observations. Hindrances for producing such diagrams have been the difficulty in properly orienting the images, as well as difficulties in accurately processing them. The latter has been overcome over the last few years with the development of methods for the high quality processing of photographic full disk data \citep{Chatzistergos2022,chatzistergos_analysis_2023}. The situation is worse for the image orientation. Previous studies with KoSO data utilized markings on the plates (examples are shown in Fig. \ref{fig1:context}) to orient the images whether manually clicking on them with the computer mouse \citep{Priyal2014} or having an automatic process to identify their locations \citep{Chatterjee2016}. However, relying on such markings introduces significant uncertainty and bias in the determined orientation, due to potential mistakes in the placement of the markings on the plates, and also in the identification of their location in the digital images. Furthermore, not all of the observations include such markings. In particular, the earliest data in the archive, that is before 1907, were not complemented with pole markings at all and thus have not been incorporated in any butterfly diagram yet.
Achieving an accurate orientation of the KoSO data is particularly important for any investigation extending back to those times.

To further facilitate the exploitation of the KoSO data to their full potential, in this study we present an automatic and versatile method we developed to orient the observations without the need to rely on the pole markings. We applied the method to the data accurately processed with the techniques developed by \citep{Chatzistergos2018,Chatzistergos2019}, and produced a butterfly diagram of plage areas from the entire data collection of KoSO, covering 1904\,--\,2007. In Section \ref{s:data}, we will briefly describe the data and the processing we applied on them. In Section \ref{results} we present the updated plage latitude vs. time diagram (i.e. plage butterfly diagram). We conclude with a summary and discussion in Section \ref{sec:conclusions}.

\section{Data and their Processing}
\label{s:data}
The main features of the KoSO Ca II K observations and methods employed to digitize, process and calibrate them are described in several papers, so that we restrict ourselves presenting a summary only.

\subsection{KoSO Photographic Archive}
\label{subs:photo}

KoSO is among the first observatories to start regular multi-wavelength surveys of the Sun at the beginning of the 20th century \citep{Hasan2010}, in white light and in the \ca\ line since 1904, and in the H$\alpha$ line since 1912. A Hale spectroheliograph, which produces a 30~mm image of the Sun with the help of a 30~cm Cooke photo-visual triplet, is used to take the \ca\ observations by scanning the disk of the Sun. A Foucault siderostat, containing a single mirror of 46~cm diameter, reflects the beam of light on a 30~cm lens, which is ultimately fed to a spectroheliograph to obtain the observations \citep{Bappu1967}. This spectroheliograph consists of two prisms which disperse the light, and then a Lyot filter is used to ensure that only a narrow spectral region centered at 393.367~nm wavelength with a pass-band of 0.05~nm passes through it \citep{Bappu1967, Priyal2017}. These observations were recorded on photographic glass plates until February 1989, while after that they were taken on photographic films (in the following we will refer to both glass plates and film as plates). There were also a few instances before February 1989 when observations were stored on films. The plates were developed in a pitch-dark room and then they were placed inside an envelope and stored in a humidity-free environment. The date and time of observations (\tobs) were written on the margin of the plates as well as on the envelope before archiving/preservation.

\begin{figure}[!htbp]
\centerline{\includegraphics[width=\textwidth,clip=]{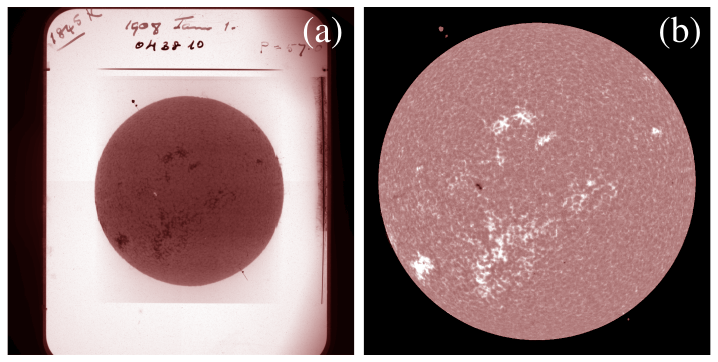}}
\caption{(a) Representative image of the photographic plate with the raw negative observation taken on 1st January 1908 at 10:08 IST (04:38 UTC) from the KoSO \ca\ digital archive, (b) the calibrated image of the corresponding observation derived from the careful calibration and processing of the raw negative data. The calibrated image is shown in the range of [-0.4, 0.4] in contrast values, while the raw image is shown to its full range of values. Note the markings to denote the poles of the Sun, two circles for the North pole and one small circle for the South pole.}
\label{fig1:context}
 \end{figure}

\subsection{Digitization}

To allow the larger community to utilize this extensive dataset, it was essential to transform it into digital format. Two such large-scale digitizations have been performed for the KoSO archive \citep[see][]{Chatzistergos2019b}. The first such attempt was restricted to only a sample of the data covering 1907\,--\,1999 \citep{Makarov2004}. 
The digitization was performed with a linear array of 900~pixel and the images were stored in 8-bit JPG file format \citep{Makarov2004}. 
However, some inhomogeneities were reported in the digitised KoSO data \citep[see][]{Chatzistergos2019b}. Furthermore, an important part of the data collection, that is the data over 1904--1907 and 2000--2007, were completely missing. This led to an effort to re-digitise the KoSO data.  The second digitization was performed with a better scanning device than the one used for the first digitisation and it aimed at including the entire data collection covering 1904\,--\,2007 \citep{Priyal2014, Chatterjee2016}. The digitization setup consisted of a 1\,m uniform source sphere with a 35\,cm opening. The source sphere contained 3\,--\,4 current control light sources to provide highly uniform lighting. At the opening of the sphere, a plate holder with a horizontal slider was used to hold the plates. A cryogenic-cooled (at -100$^\circ$~C) 4k\,$\times$\,4k CCD camera with a 16-bit readout, attached to a vertical slider, was used for digitizing the \ca\ observations taken on photographic plates. The same setup with a film holder instead of a plate holder is used for digitizing the observations taken on photographic films. After digitization, these images were stored in 16-bit FITS \citep[Flexible Image Transport System;][]{wells_fits_1981} format, with a pixel scale of $\approx$\,0.9\,arcsec/pixel (an example image is shown in \autoref{fig1:context}a). A set of five flat-field images are taken for each month of data after digitization for that particular month, whereas the dark is subtracted during the digitization itself \citep{Priyal2014}. Finally, the filename of FITS  images is given based on the date and time of observation (\tobs), noted from the plate or envelope carrying these plates, which is required for using the data.

The above digitization setup and procedure are now permanently available at KoSO for digitization of photographic observations. Indeed, recently, we found some inconsistencies in the digitized data for six years  (1957\,--\,1962), and we also recovered a few more plates in the archive missed by both previous digitization drives, mostly between 1957 and 1961. Hence, we have re-digitized all the data for this period with the same setup as used for the latest digitization, in order to remove any data in-homogeneity and incorporate the newly-found plates. In this work, we use the data from the second digitization \citep{Priyal2014}, along with the re-digitized data over (1957\,--\,1962). In total there are 51773 images, which is 29615 more images than analyzed by \citet{Makarov2004, Tlatov2009, Ermolli2009} based on the first digitization of the KoSO \ca\ archive, and 2845 more images than analyzed by \citet{Priyal2014, Chatterjee2016} from the second digitization of the KoSO \ca\ archive.

\subsection{Calibration}
We used the KoSO images processed by
\citet{Chatzistergos2019b, Chatzistergos2020} along with newly digitised images over 1957\,--\,1962, which were processed with exactly the same techniques as the rest of the data \citep[see][ for details]{Chatzistergos2018}. The processing comprised three main steps: (i) pre-processing; (ii) photometric calibration; and (iii) compensation for limb-darkening and image artifacts. The pre-processing includes the flat-fielding of the CCD-camera used for the digitisation \citep{Chatzistergos2019b}, conversion of transparency (negative) to density (positive) values, and the identification of the solar disk. The latter was achieved by applying a Sobel filter and using a threshold to isolate the rim of the solar disk \citep{Chatzistergos2020}. An ellipse was fitted to the isolated rim to acquire the values of the center of the ellipse and the semi-major and -minor axes. These were then used to resample the images and return circular solar disks \citep{Chatzistergos2020,Chatzistergos2020b}. Such corrections were required due to image distortions introduced  during the observation \citep[see][]{Chatzistergos2020,Chatzistergos2020b}. 

The photometric calibration is required because the emulsions used to store the photographs have a non-linear response to the incident light. The approach applied to recover the information needed for the photometric calibration is to relate the density centre-to-limb variation (CLV) of the darker parts of the quiet Sun (QS) regions measured in the photographic data to intensity QS CLV from modern CCD-based observations \citep{Chatzistergos2018, Chatzistergos2019}. In this way we construct a calibration curve for each image separately. Finally, compensation for the limb darkening \citep{Chatzistergos2018, Chatzistergos2019} allows to create contrast images (\autoref{fig1:context}b) of the form $C_{i,j}=(I_{i,j}-I_{i,j}^{\mathrm{QS}})/I_{i,j}^{\mathrm{QS}}$, where $C_{i,j}$, $I_{i,j}$, and $I_{i,j}^{\mathrm{QS}}$ are the contrast relative to the quiet Sun, intensity, and intensity of the quiet Sun for pixels $(i,j)$. The map of $I_{i,j}^{\mathrm{QS}}$ was computed with an iterative approach, involving a 2D running-window median filter as well as polynomial fitting along rows, columns, and radial segments of the solar disk after removing the active regions \citep{Chatzistergos2018}. This step allows for the removal of large scale image artifacts due to the methods  applied for the acquisition and storage of the observation. The superior accuracy of the above processing method compared to others from the literature has been extensively discussed by \citet{Chatzistergos2018, Chatzistergos2019,Chatzistergos2020,chatzistergosmodelling2020,Chatzistergos2022} and \citet{chatzistergos_analysis_2023}.

\subsection{Pole Angle Calculation}
\label{s:pole_angle}
The accurate orientation of KoSO \ca\ observations results from application of a few processing steps. First, we describe how we calculate the pole angle (orientation), i.e. the direction of solar North with respect to the vertical axis of a digitized image for a given date and time of observation (\tobs). Assuming that the siderostat at KoSO reflects the sunlight perfectly in the geographical South direction along the meridian, the rotation, $\alpha$, of the \ac{fov} caused by the siderostat is given by \citep{Cornu1900}:
\begin{equation}
    \alpha = 2\tan^{-1}{\left[ K\tan{\left( \frac{{\rm HA}}{2}\right)}\right]},
    \label{eq1}
\end{equation}
where
    $$ K = \frac{\sin{\left( \frac{L-\delta}{2}\right)}}{\sin{\left( \frac{L+\delta}{2}\right)}}.$$
In \autoref{eq1}, {\rm HA} is the hour angle of the Sun \citep{Smart1977}, $L$ is the latitude of the observatory, and $\delta$ is the pole angle (\ang{90} - {\rm declination}) of the Sun at \tobs. To calculate hour angle of the Sun we have used a IDL routine in SolarSoftWare \citep[SSW;][]{freeland_data_1998} {\it sunpos.pro} and {\it eq2hor.pro}. We use the following co-ordinate of the observatory Latitude($L$)=\ang{10.23} (\ang{10;13;50})N and Longitude = \ang{77.47} (\ang{77;28;07})E (\url{https://www.iiap.res.in/kodai_location}). The orientation of digitized images is in such a way that, to get the solar North at the top of the images we need to rotate them by the complementary angle of $\alpha$. 
Therefore, the pole angle $(\Theta)$ to bring the solar North to the top of the image is given by
\begin{equation}
    \Theta =\ang{90}-\alpha.
    \label{eq2}
\end{equation}
It is obvious from Equations~\ref{eq1} and \ref{eq2} that, getting $\Theta$ is straightforward, provided we have the correct \tobs.

Prior to the use of \autoref{eq1} and \ref{eq2} for the calculation of $\Theta$, we compared $\Theta$ calculated with the method of \cite{Priyal2014}. In most cases, the difference amounts to a few degrees $(\pm\ang{2})$, but for $\approx 0.5\%$ of the data the differences are substantial $(>\ang{10})$.  Manual inspection of these cases proved that our approach returned the correct orientation, while the pole markings were erroneously placed on the plates. A more detailed discussion of the comparison can be found in \autoref{previous_method}. Note that the above comparison is based on the assumption that all the \tobs\ is correct, which is not the case as we discuss in the following sections.

\subsection{Inconsistencies in time of observations}
\label{sec:correctingtime}
The approach described above allows to get an accurate $\Theta$ if \tobs\ is known precisely. However, some inconsistencies in the date and time of observations of the digitized images have previously been reported \citep[e.g.][]{Chatzistergos2019b, jha2022thesis}, which can affect the image orientation as well as results from any analysis with inexactly oriented data. 
For instance, this has the potential implication of previously derived plage areas including values that have been assigned to the wrong time periods. It is thus worth trying to tackle those inconsistencies \citep[see][]{jha2022thesis}.

\begin{figure}[!htbp]
\centerline{\includegraphics[width=\textwidth,clip=]{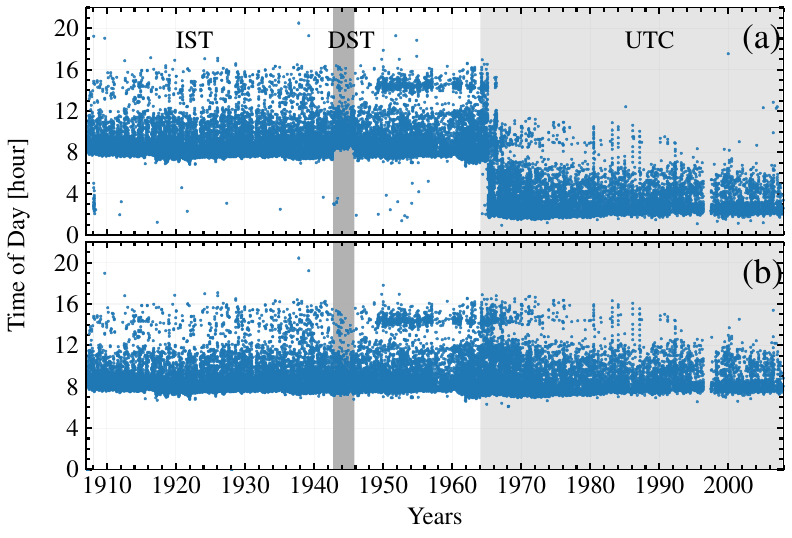}}
\caption{ (a) Time of observation at KoSO over the period of observations. Different background shadings are used to mark the different time conventions adopted, IST (+05:30 UTC; white), UTC (light grey) and DST (+06:30 UTC; dark grey), (b) same as panel (a) after applying the corrections of time of observations described in Section~ \ref{sec:correctingtime}.}
\label{fig3:time_obs}
 \end{figure}

We note that, there are systematic inconsistencies in \tobs~of the files as well as random mistakes. Systematic inconsistencies arose due to conversion from \ac{utc} to \ac{ist}, because of either an incorrect conversion or a change of the convention used when writing the data on the plate/folder. Unfortunately, there are also random mistakes in the date and time of observation, which could have arisen at any stage, either when the date/time was written on the plate/folder or when the plate was digitized and the time was used to name the file. To illustrate these inconsistencies, in \autoref{fig3:time_obs}a we show the time of observation, extracted from the filenames, as a function of the observational date. In \autoref{fig3:time_obs}a we notice the following:
\begin{enumerate}
    \item Before February 1964 time is recorded predominantly in \ac{ist} (+05:30 \ac{utc}), while \ac{utc} was used afterwards. However, we found that in the years 1964 (February on-wards) and 1966 (mostly in the first half),  97\% and 43\% of data, respectively, have been converted to \ac{ist} from \ac{utc}, during the digitization itself.
    
    \item There is a systematic shift of one hour in the time of observation, from September 1, 1942 to October 15, 1945 due to the Daylight Saving Time (DST; +06:30 UTC)\footnote{\url{https://web.archive.org/web/20230517151641/https://www.timeanddate.com/time/change/india/kolkata}} followed in India.
    
    \item From January 1 to January 24, 1908, most of the \tobs\ is in UTC (timestamp in \autoref{fig1:context} is in \ac{utc}), with few exceptions which are in \ac{ist}. 
    
    \item Observation times beyond the usual daylight period (06:30 to 17:00), which are primarily due to mistakes in naming of files during digitization or confusion between IST and UTC.
\end{enumerate}

We have accounted for the above inconsistencies between IST and UTC. However, as we previously mentioned, there are also mistakes in the dates/times of the observations which might have gone unnoticed. Furthermore, we cannot rule out that some of the isolated outliers in \autoref{fig3:time_obs}a might have been due to mistakes in writing the date/time and not due to conversion between IST and UTC of the series. We note that the data in \autoref{fig3:time_obs} only cover the period 1907\,--\,2007, i.e. they lack the first 3 years of KoSO \ca\ observations, which are short of information on the \tobs.

We correct the inconsistencies in the \tobs\ in two strides, each consisting of several steps. These are briefly discussed in the following, while more details are given in Appendix \ref{compare_cont}. In an attempt to bring all \tobs\ values to the same time zone, and since most of the timestamps are in IST, we decided to go with IST. We stress that this refers to the time stored in the header of the FITS files, the time is converted to UTC for all images when computing the butterfly diagram. As the first stride of correction we considered accounting for the major inconsistencies in \tobs\, as they were pointed out earlier. In the second stride, we have attempted to identify and correct residual mistakes in the \tobs\ following \citet[their Section~2.3]{jha2022thesis}. In \autoref{fig3:time_obs}b we show the corrected \tobs\ as a function of date of observation. Additionally, we also note that there is gradual change in \tobs\ post 1960 which is an attribute of the early observation in the morning at KoSO, but the reason behind the start of early observation is unknown to us.

\begin{figure}[!htbp]
\centerline{\includegraphics[width=\textwidth,clip=]{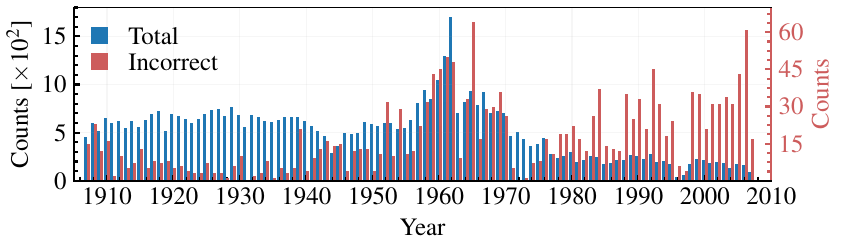}}
\caption{Number of \ca images per year in the KoSO archive (blue) along with the number of images that we estimate to have incorrect \tobs\ even after our reduction to same time convention (red and ordinate axis on the right-hand side; see Section \ref{sec:correctingtime}).}
\label{fig5:distribution_year} 
\end{figure}

After applying all these corrections, we found that $\approx3.4\%$ of observations still remain with potentially incorrect \tobs (timestamps). The most probable reason for the incorrect \tobs\ of these observations is due to the mistake in noting \tobs\ during the observations. In \autoref{fig5:distribution_year} we show the yearly counts of images with potentially incorrect \tobs\ along with total number of observations. Note the different scales used for the two counts in \autoref{fig5:distribution_year}. \autoref{fig5:distribution_year} shows an increase of the number of images with incorrect \tobs\ over the periods 1957\,--\,1970 and 1975\,--\,2007, which cover Solar Cycles 19\,--\,20 and the most recent decades, thus potentially affecting the behavior of the envelope and long-term evolution of the solar activity derived from analysis of KoSO \ca\ observations.

\subsection{Recovering the image orientation over 1904\,--\,1906}

We found that approx 3.4\% of the data suffer from potentially incorrect \tobs\ after conversion of the \tobs\ archived with the observation to the same time convention. Besides, we also note that observations taken before 1907 do not carry a time of observation at all (only the date is available). Furthermore, we observe that the digitized images derived from the KoSO archive may suffer from further orientation issues due to, e.g., how exactly the plate was placed in the digitization device.

To overcome all the above problems, we developed a cross-correlation technique to determine the orientation of the KoSO \ca\ observations. The method tries all possible orientations between two subsequent images and selects the one that minimizes the sum of squared residuals of their contrast values as the correct one. This is done in two iterations. First, all angles are checked with a step of \ang{1}, while the second iteration reduces the step to \ang{0.1} and restricts the search of the angle within $\pm$\ang{1} of the optimum value returned during the first iteration. We use the calibrated and limb-compensated images to remove as much as possible the influence of limb darkening and image artifacts, which would have affected the results from  the cross-correlation of the images. The images were then resized to have the same dimensions (we note that this was done only for the estimation of the rotation, otherwise the original file size was maintained). We also applied differential rotation to rotate every image in question  to \tobs\ of the corresponding image acting as the reference (the one with known orientation) to match their observing times. This is done using \textit{drot\_map.pro}\footnote{See \url{https://hesperia.gsfc.nasa.gov/ssw/gen/idl/maps/drot_map.pro} for details about the routine.} of SSW.
To work correctly, this in turn requires that the reference image of each comparison is properly oriented. For orienting the images we used the observation available in the archive that is closest in time and has been already oriented. We restricted the search for such a fiducial image to within a 3-day interval around the time the image to be rotated was recorded. When there were no data available within the desired limits, we used observations from the Meudon dataset \citep[as analysed by][]{Chatzistergos2020} to act as reference.

We used the above cross-correlation approach on all data before 1907 to determine the pole (rotation) angle. However, we also used the cross-correlation approach on the data after 1907 in order to validate our pole angle calculation based on Equations \ref{eq1} and \ref{eq2} and to identify any images with wrong timing or possible misalignment of the plate during the digitization. To distinguish between potentially correct and wrong orientations we imposed a threshold of \ang{5} between the rotation angle determined with Equations \ref{eq1} and \ref{eq2} and the one from the cross-correlation approach. If the difference between the two angles was greater than the threshold, or if the cross-correlation returned a different flipping (e.g., East-West Flipped in the cases where images are scanned from the wrong side), then we checked the observation date in the raw images to identify if there was a mistake in the date/time of the digital file. However, there were still a few instances, where we were unable to correct the observational date and time of the images (e.g. if there was no date information written within the scanned area of the plates) or there was still disagreement between the rotation angle derived with the cross-correlation and the one with Equations \ref{eq1} and \ref{eq2} (suggesting that there might have been a mistake in writing the date/time on the plate). In these cases we simply stored the rotation angle determined from the cross-correlation in the FITS files header and used this one instead to align the images appropriately.

\subsection{Segmentation}
\label{sec:segmentation}

The reliable orientation of the KoSO \ca\ observations accurately processed with the methods developed by \cite{Chatzistergos2018} allows for investigations of plage regions going beyond those that could be achieved in previous studies by e.g. \cite{Chatzistergos2020}. In this light, we also segmented the oriented data following \cite{Chatzistergos2019}. In brief, plage regions were identified with a contrast threshold, which was taken as a multiplicative factor to the standard deviation of the quietest parts of the disk. The level of the quiet Sun regions was identified with a variant of the method by \cite{ribesfractal1996}. This thresholding was used to produce binary maps of the solar disk where the plage regions have the value of one and all other locations the value of 0. We used these masks to compute/derive the butterfly diagram of the daily plage areas. We considered plage areas in millionths of solar disk fraction ($\mu$DF) within latitudinal strips of \ang{1}, similarly to \citep{harvey_cyclic_1992}.  For days when multiple observations exist, we took the average plage area within each latitude bin for all images of that day.

\section{Results}
\label{results}

\subsection{Plage Butterfly Diagram}
\label{butterfly}

\begin{figure}[!htbp]
\centerline{\includegraphics[width=\textwidth,clip=]{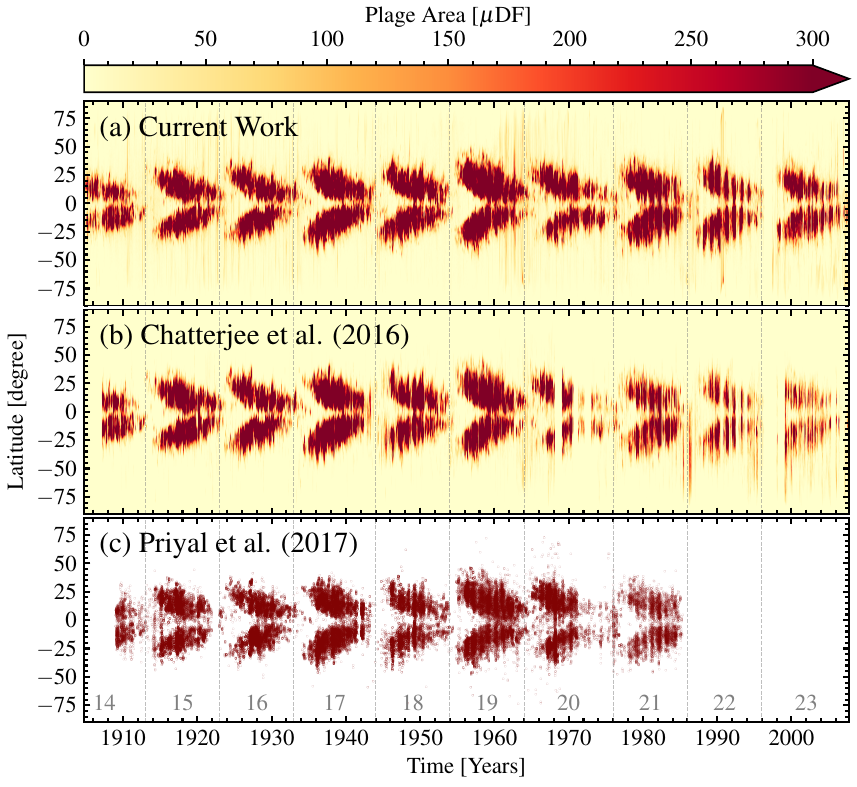}}
\caption{The plage area millionths of solar disk fraction ($\mu$DF) for each latitude bin of \ang{1} for each observation, averaged over all the data of the day in case of multiple observations, as a function of time, (a) for current work, (b) generated using the plage regions identified by \citet{Chatterjee2016} and (c) from \citet{Priyal2017}. However, we note that the butterfly diagram with data by \citet{Priyal2017} shows the latitude of the centroid of individual plage regions and not latitudinally averaged areas.}
\label{fig4:butterfly} 
\end{figure}

\autoref{fig4:butterfly}a shows the butterfly diagram of plage areas derived from our processing of the KoSO \ca\ observations. It is evident from this figure that similar to sunspots, plage also appear at higher latitudes at the beginning of the solar cycle (SC), and as the SC progresses, they move towards the equator. This behavior is in good agreement with earlier findings of \citet{harvey_cyclic_1992, Chatterjee2016} and \citet{Priyal2017}.

To compare our findings with earlier results from the literature, we have also plotted in \autoref{fig4:butterfly}b and \ref{fig4:butterfly}c the butterfly diagrams of KoSO plage areas produced by \citet{Chatterjee2016} and \citet{Priyal2017}, respectively. Both of these studies created their butterfly diagram by defining the centroid latitude of each connected plage region. To make the comparison to our butterfly diagram (which plots areas in latitudinal bins irrespectively of connectivity of plage regions), we use the identified plage region masks provided by \citet{Chatterjee2016} to create a butterfly diagram in the same way as in this work, i.e., disk fractional area in each latitudinal bin of \ang{1}. However, \citet{Priyal2017} only provides the mean latitude and area for individual plage regions along with the time of observation. Therefore, in \autoref{fig4:butterfly}c, we plot the butterfly diagram as it is presented in their paper. \autoref{fig4:butterfly} clearly shows the butterfly diagram derived from our processing includes more data (\autoref{fig4:butterfly}a), by covering the entire period of operations of the KoSO and having fewer data gaps, than both the other compared diagrams (\autoref{fig4:butterfly}b and \ref{fig4:butterfly}c). Both \citet[]{Chatterjee2016} and \citet{Priyal2017} ignored many images (mostly after 1960, while \citealt{Priyal2017} also ignored all data after 1985) due to their degrading quality, while no data exclusion was performed in our case. Comparison between \autoref{fig4:butterfly}a and \ref{fig4:butterfly}b also suggests that our processing improves the accuracy of the data representative of the minima at approx 1985 and 1995,  i.e. minima SC 21--22, SC 22--23. This could be important when analyzing long-term trends of solar processes.

This further suggests that images taken around 1961 and 1990 still suffer from some inaccuracies and the stripes we see in \autoref{fig4:butterfly}a are due to the presence of artifacts in the digitized images, which are falsely identified as plage regions \citep{Chatzistergos2019}. These stripes are not present in \autoref{fig4:butterfly}b simply because those data were not considered in the analysis of \citet{Chatterjee2016} at all. Moreover, our results also appear to be more consistent than the others for the periods 1942--1945 and 1986 in terms of the distribution of plage over latitude. This is due to the correction of the timing of observations applied by us (see Sect. \ref{sec:correctingtime}).

\subsection{Latitudinal Extent of the Plage}
\label{lat_extent}

\begin{figure}[!htbp]
\centerline{\includegraphics[width=\textwidth,clip=]{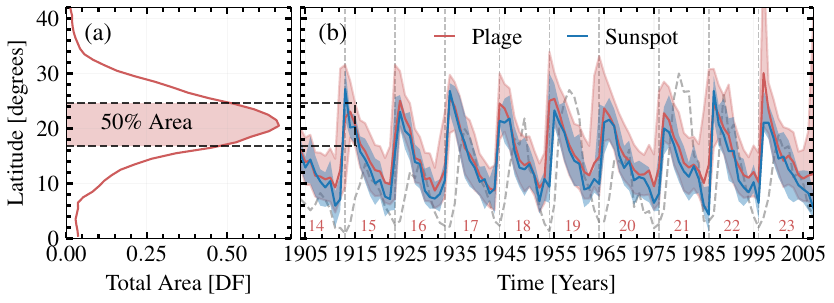}}
\caption{(a) Normalized latitudinal distribution of plage areas averaged over the year 1915 after combining both hemispheres. The black dashed horizontal lines and red shading mark the bands between 25\% and 75\% percentile of the area distribution. (b) The yearly variation of median 
latitude of KoSO plage areas (solid red line) along with the latitude band that includes 50\% of each year's \ca\ plage areas (red shaded surface). Also shown is the latitudinal extent of KoSO sunspot areas by \citet[][blue curve for annual median values and blue shaded surface for the band covering 50\% of the areas]{Jha2022}. The Grey dashed line represents the scaled annual mean plage area and is over plotted to show the progression of the SC. The \bkj{black dashed} rectangle
marks the latitudinal band including 50\% of plage areas for 1915, the latitudinal distribution of which is shown in panel (a). The numbers at the bottom of the panel denote the conventional SC numbers.}
\label{fig5:lat_ext} 
\end{figure}

We used the derived butterfly diagram to investigate the latitudinal extent of plage regions with respect to that of sunspots. 
For sunspot areas we use the recently updated and extended KoSO sunspot area series by \citet{Jha2022}, so that we compare chromospheric plage and photospheric sunspot data obtained under nearly the same observing conditions. We calculate the yearly latitude band in which 50\% of plage/sunspot area lies. To do this, we combined the plage/sunspot areas in both hemispheres. Then we summed up all the plage/sunspot areas within each \ang{1} latitude bin over a year (see \autoref{fig5:lat_ext}a for the plage observed during the year 1915 as an example). The central latitude band in which 50\% of the plage area for the selected year is covered is represented by a shaded region in \autoref{fig5:lat_ext}a.  In \autoref{fig5:lat_ext}b, we plot this latitude band for plage regions as a function of time along with the similarly determined latitude band for sunspots. We calculate the area of plage regions in each latitude bin (\ang{1}), while for sunspots the mean centroid location of a sunspot is used as the latitude and the areas of all sunspots whose centroids lie within a particular latitude bin are assigned to that bin. An example of a study making use of the latitudinal distribution of \ca\ plage is given by \cite{Hofer2024}.

In \autoref{fig5:lat_ext}, we observe that the  median value of the latitude of plage regions and the upper end (75 percentile) of the latitude band covered by them are almost always higher than those for sunspots, whereas the lower limit (25 percentile) remains close to that of sunspots. Additionally, we found that the mean latitude of plage areas calculated over the entire period is $20.5\%\pm2.0$ higher than that of sunspots.
This behaviour is consistently present throughout the observation period, independently of the phase of the SC. 
Over a comparatively short period of time during late 1990s, plage areas extend to particularly high latitudes. This is unfortunately due to residual artefacts due to the lower quality of the data over this period \citep{Chatzistergos2019}.

\subsection{Hemispheric Asymmetry}
\label{assy}
\begin{figure}[!htbp]
\centerline{\includegraphics[width=\textwidth,clip=]{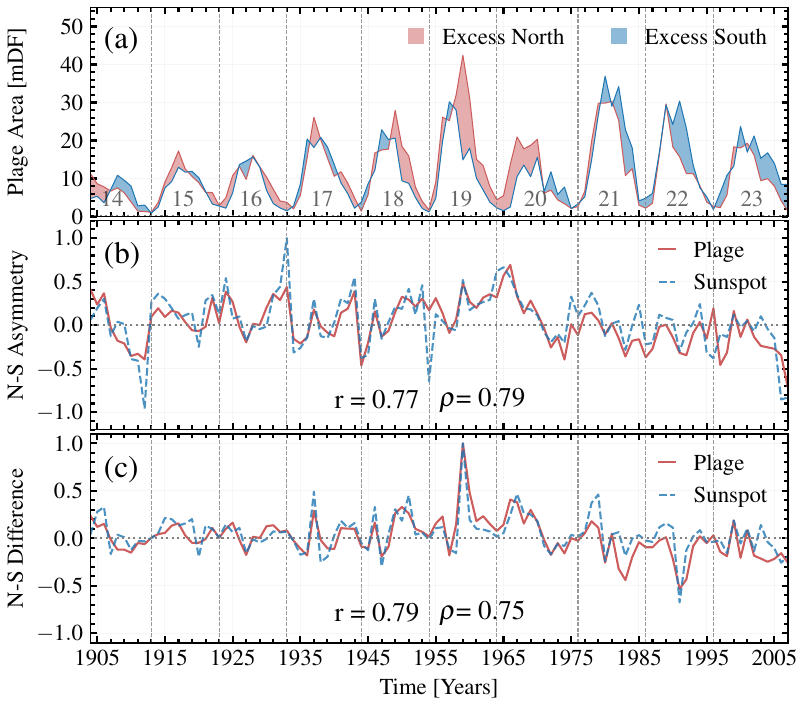}}
\caption{(a) Hemispheric yearly averaged plage area as a function of time. The red (blue) shaded regions represent the period where the northern (southern) hemisphere dominates over the other hemisphere. (b) The N-S asymmetry calculated using \autoref{eq3} for plage and sunspot areas and the Pearson ($r$) and Spearman ($\rho$) correlation coefficients are listed at the bottom of the panel whereas (c) shows the normalized difference (with their respective maximum values) of the yearly averaged plage and sunspot area in two hemispheres.}
\label{fig6:assymetry} 
\end{figure}

The Sun shows asymmetry in the magnetic activity manifested in its northern and southern hemispheres. This asymmetry of the Sun (N-S asymmetry, hereafter) has been studied using various indices of solar activity, e.g., sunspot number \citep{Veronig2021}, sunspot area \citep{Vizoso1990}, \ca brightness \citep{Bertello2020}, and magnetic flux \citep{Hathaway2016}. Here, we utilize the butterfly diagram derived from our processing of the KoSO \ca\ observations to study the N-S asymmetry in chromospheric plage areas. In \autoref{fig6:assymetry}a, we plot the yearly averaged area calculated separately for the northern ($A_{\rm North}$) and southern hemispheres ($A_{\rm South}$). In \autoref{fig6:assymetry}b, we also show, as a function of time, the N-S asymmetry calculated as:
\begin{equation}
    \text{N-S Asymmetry} = \frac{A_{\rm North}-A_{\rm South}}{A_{\rm North}+A_{\rm South}}.
    \label{eq3}
\end{equation}
To further investigate the relationship between plage and sunspots, we computed the correlation coefficients between the N-S asymmetry observed for plage and sunspots.

We notice that the southern hemisphere dominates during SCs 21\,--\,23, while the northern hemisphere dominates from the declining phase of SC 18 until the maximum of SC 20. Earlier SCs appear more balanced between the two hemispheres although for SCs 14 and 18, each hemisphere dominates roughly half of the cycle.

Interestingly, the rise of SC 19 seems to be nearly balanced in the two hemispheres until the peak is reached in the southern hemisphere. Afterwards the plage areas in the northern hemisphere continue rising for another two years, and then the northern hemisphere dominates until the declining phase of SC 20.
We also notice that during the descending phase of SC 20 and the following minimum, when we previously (Section~\ref{butterfly}) reported residual artifacts at high latitudes, there is an excess of plage areas in the southern hemisphere. During SCs 16, 17, and 21\,--\,23 one hemisphere has double peaks, while the other hemisphere shows a single peak, which is in agreement with the earlier work by \citet{Chowdhury2022} for SCs 16 and 17, whereas they have not used the data after 1980 to discuss SCs 21\,--\,23.

To compare our results for the N-S asymmetry of plage areas to that in sunspot areas, we again used the KoSO white-light sunspot area series by \citet{Jha2022}. We calculated the asymmetry in sunspot area in the same way as we did for the plage areas, and both are shown in \autoref{fig6:assymetry}b. The two series follow each other quite closely. The main differences we notice is the slightly more balanced sunspot areas over the two hemispheres during SCs 21\,--\,23 compared to plage areas. Multiple sharp excursions in the sunspot area asymmetry are seen near activity minima. These are due to the fact that during minima sunspot areas drop to very low values close to 0, while plage areas are always higher than sunspot ones. Nonetheless, the small plage coverages around activity minima likely also lead to (less dramatically) excessive asymmetries in the plage areas at such times. 

To avoid such excursions during times of low coverage, in \autoref{fig6:assymetry}c, we display the difference of the yearly averaged plage and sunspot area coverage of the north and south hemispheres (N-S Difference $=A_{\rm North}-A_{\rm South}$). To keep the values within the same vertical scale as in \autoref{fig6:assymetry}b, we plot the N-S Difference after normalizing it by the maximum of the respective differences (27.5 mDF and 1031 $\mu$Hem, respectively for plage and sunspot). It is evident from \autoref{fig6:assymetry}c that the large values of N-S asymmetry during the SC minima are an artifact of poor statistics. It is interesting to note that the pronounced and comparatively long-lasting dominance of the northern hemisphere during the second half of SC 19 and the first half of SC 20 is seen consistently in both \ca\ plage and sunspot areas.  The Pearson ($r$) and Spearman ($\rho$) correlation coefficients for N-S Asymmetry with its sunspot counterpart are $r=0.77$ and $\rho=0.79$. The respective coefficients for N-S Difference and its sunspot counterpart are $r=0.79$ and  $\rho=0.75$), respectively.

\section{Summary and conclusions}
\label{sec:conclusions}

KoSO has provided a wealth of solar observations since the beginning of the 20th century in various wavelength intervals, most prominently in the \ca line. That being said, these data have not been utilized to their full potential due to inconsistencies in the  orientation of the images. This limitation particularly affects studies that require precise locations of features on the Sun \citep[e.g.][]{mishra_ca_2024}, e.g, latitudinal distribution of plage \citep[e.g.][]{Hofer2024}, N-S asymmetry of plage areas, etc. Therefore, in this study we developed and applied a new automated process to orient the KoSO \ca\ observations. Our approach is more robust than previous methods applied to KoSO data, which relied on manually placed pole markings on the drawings and subsequently manual clicking on the markings to determine the orientation of the observations. Furthermore, we identified the difference between our computed rotation angle to previous estimates to be $\pm2^\circ$ assuming there is no inconsistencies in Time of Observation (\tobs). However, $\approx$0.5\% ($\approx200$)

of the images were found to have pole markings at wrong positions, resulting in an error in the rotation angle that is greater than 10$^\circ$. Additionally, we also corrected the time of observation, \tobs, for $\approx10\%$ of the data which suffer from incorrect \tobs, again leading to an error in the rotation angle. The newly developed method for image orientation, and knowledge of issues with \tobs, worked out here for \ca\ observation will be extended for application to KoSO H$\alpha$ and \ca\ prominence observation as all of them are taken with the same instrumental setup. Furthermore, it will also be useful for observations from other archives taken with a similar instrumental setup.

In our analysis we further benefited from more accurate processing to compensate for artifacts affecting the images and to perform the photometric calibration than in previous studies requiring an accurate image orientation. Based on this, we produced the first complete butterfly diagram from KoSO \ca\ data, which covers the entire period of observations between 1904\,--\,2007.

We have also compared our butterfly diagram to those previously published in the literature. We showed that due to our more accurate processing we are able to produce a butterfly diagram including more data than previously possible. To remove some inconsistencies we corrected issues with the dates and times of the observations that existed in previously published butterfly diagrams of plage areas from KoSO \ca\ data. We then used the obtained butterfly diagram to compare the latitudinal extend of plage areas to that of sunspots. We found that the mean latitude of plage areas is higher than that of sunspots irrespective of the phase or the strength of the SC. 
We expect that this behavior has to do with Joy's law \citep{Hale1919} and the fact that plage are more strongly concentrated in the following polarities of active regions, which are on average further from the equator than the leading polarity (which typically harbors the larger spots). The longer lifetime of plage also means that they are more affected by diffusion and are also more likely to be dragged by the meridional flow towards the poles than sunspots are.

Finally we also investigated the N-S Asymmetry in plage areas.
We found that the northern hemisphere dominated plage areas during SCs 19 and 20, while the southern hemisphere dominated between SCs 21 and 23. This consistent dominance of the southern hemisphere could be an artifact of degraded data quality, but it can also be linked to the long-term variation of the solar magnetic equator as suggested by \citet{Pulkkinen1999}. However, further investigation is needed to substantiate such claims.
We also compared the N-S Asymmetry of plage areas to that displayed by sunspot areas based on white-light observations obtained at KoSO under the same observing conditions. We found that the asymmetry in plage areas is  well correlated ($r=0.77, ~\rho=0.79$) with the asymmetry observed in the sunspot areas. 

With this study we extended the usability of KoSO \ca\ data, forming one of the most prominent such historical archives, for wider scientific applications. Therefore, corrected data and obtained results will contribute to achieving improved reconstructions of past solar surface magnetism \citep[e.g.][and references therein]{Chatzistergos2019c,Chatzistergos2022, Mordvinov2020,Hofer2024} and irradiance variations \citep[e.g.][]{Chatzistergos2021,chatzistergos_long-term_2023} from \ca observations. Information derived from the century-long \ca\ plage butterfly diagram obtained here has the potential to provide valuable inputs to dynamo models and understanding of SC evolution. Besides, the KoSO \ca\ data accurately processed and oriented derived from this study will allow for further research of the properties of the solar rotation and large-scale flows over roughly a century. First results from such studies have been recently presented by \citet{mishra_differential_2023} and more will follow. More importantly, we produced a dataset of KoSO \ca\ accurately oriented observations along with the data for corrected \tobs\ and image orientation will be made available to the public for use in further investigations.

%
\appendix

\section{Comparison of pole angle calculation methods}
\label{previous_method}
\begin{figure}[!htbp]
\centerline{\includegraphics[width=\textwidth,clip=]{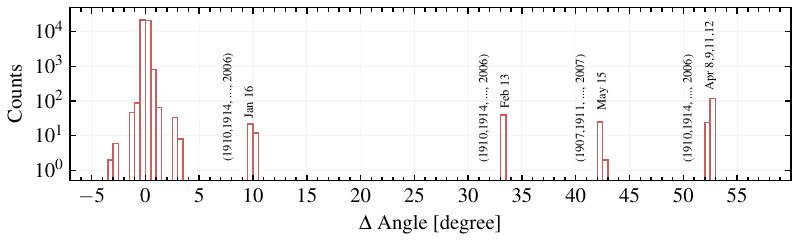}}
\caption{Distribution of differences in rotation angle ($\Delta \Theta$) calculated from the \bkj{lookup table provided at KoSO} and our estimate based on \autoref{eq2}. The peaks at $\Delta \Theta> \ang{10}$ correspond to particular dates repeating every four years. Theses dates for the first two and last occurrences in the series are written on the right side of the peaks.}
\label{fig7:ang_diff}
\end{figure}

Here we present the comparison of our new automatic method of estimating $\Theta$ from \autoref{eq1} and \ref{eq2} to the one used by \citet{Priyal2014}. These authors have calculated $\Theta$ by manually clicking on the pole markings in the images (see double and single dots right outside the solar limb in \autoref{fig1:context}, which represent the position of the North and South poles, respectively). These pole markings are drawn during the observation or just before digitization using a tabulated value (lookup table) provided at KoSO. To compare $\Theta$ calculated from these two methods, we firstly used an automatic code to extract $\Theta$ from the KoSO table  ($\Theta^\prime$ here-after) for a given \tobs\ and then used that \tobs\ to get $\Theta$  based on \autoref{eq2}.
We note however, that the rotation angle derived by \citet{Priyal2014} includes the extra uncertainty of drawing the polar markings on the photographic plates as well as manually clicking on them on the digital images with the computer mouse.
In \autoref{fig7:ang_diff}, we show the histogram of the differences ($\Delta\Theta=\Theta-\Theta^\prime-$) in the computed angles, with bin size of \ang{0.5}. \autoref{fig7:ang_diff} implies that the distribution has the most prominent peak within \ang{\pm 2}, which tells us that, in most cases ($\approx99.5\%$), our new method is in good agreement with the tabulated \citep[used by][and others]{Priyal2014} one, provided we use the same \tobs\ in both these cases, irrespective of their correctness. This analysis helped us to understand the difference between two methods for a given \tobs.

\begin{figure}[!htbp] 
\centerline{\includegraphics[width=\textwidth,clip=]{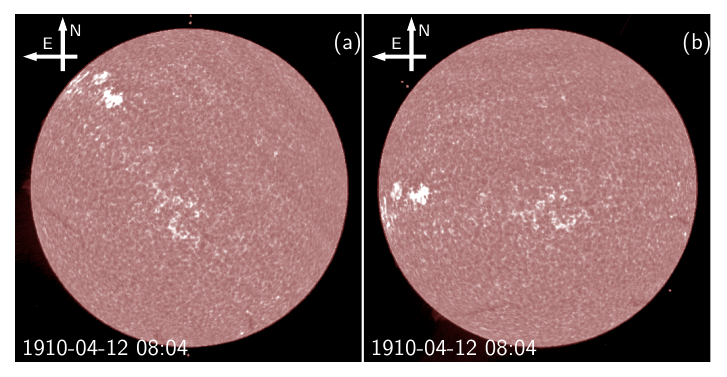}}
\caption{KoSO \ca observation taken on 12 April 1910. The images were rotated with the angle (a) tabulated by  \citet{Priyal2014} and  (b) estimated by us using Equations~\ref{eq1} and ~\ref{eq2}. The arrows at the top left part of the image denote the expected orientation of the observations. 
Note that the double dot is the North pole marking drawn by the observer on the photographic plate. These markings were used by \citet{Priyal2014} to orient the images. The images are shown after the compensation of the limb darkening and removal of large-scale artifacts, and they are displayed in the range of contrast values of [-0.4, 0.4].}
\label{fig8:app_fig1} 
\end{figure}

We note that there are cases when these methods differ significantly ($\Delta \Theta> \ang{10}$) from each other. 
The dates and times of observation (\tobs) corresponding to these peaks are restricted to a few specific dates (marked in \autoref{fig7:ang_diff}), which are repeating every four years. We visually inspected the corresponding images to check which of the methods returned the wrong rotation angle.  
We used the rotation angles from our analysis ($\Theta$) and that calculated from the KoSO table ($\Theta^\prime$), and compared the images to nearby observations. An example is shown in \autoref{fig8:app_fig1}.
We noticed that in these images although after the rotation with the angle by $\Theta^\prime$ (\autoref{fig8:app_fig1}a) the North pole markings (two small dots) appear at the top, the distribution of plage appears tilted. 
In particular in \autoref{fig8:app_fig1}a plage regions appear at extremely high latitudes, especially considering the phase of the SC of that observation (being in the descending phase of SC 14). 
The rotation with our method (\autoref{fig8:app_fig1}b) resulted in the North pole markings being at roughly off by \ang{45}, but plage regions appear at more realistic latitudes. To validate which plage region distribution is correct for these observations, we compared them with the observation nearest in time to the one under consideration. 
This allowed us to ascertain that the pole markings were placed at the wrong positions for these cases.
The rather systematic time difference of these observations (repeating every four years) suggests that this was most likely an error in the algorithm when computing the rotation angle before making the table and drawing the pole markings. This shows the robustness of our method to such mistakes. It is also worth noting that the method developed in our study can be applied to accurately derive the orientation of the KoSO observations obtained at other spectral regions than \ca\, e.g. 
H${\alpha}$ and \ca\ prominences.

\section{Manual identification of incorrect \tobs\ by comparing with nearest observations.}
\label{compare_cont}
As mentioned in the main text, the images unfortunately suffer from mistakes in the recorded observational dates and times, which can hamper the performance of our approach to orient them.
We discussed ways to mitigate some systematic issues due to such mistakes, but more isolated cases remained.
In the following we describe the steps we undertook to try to identify such cases with wrong observational date/time and account for this error.

\begin{figure}[!htbp]
\centerline{\includegraphics[width=\textwidth,clip=]{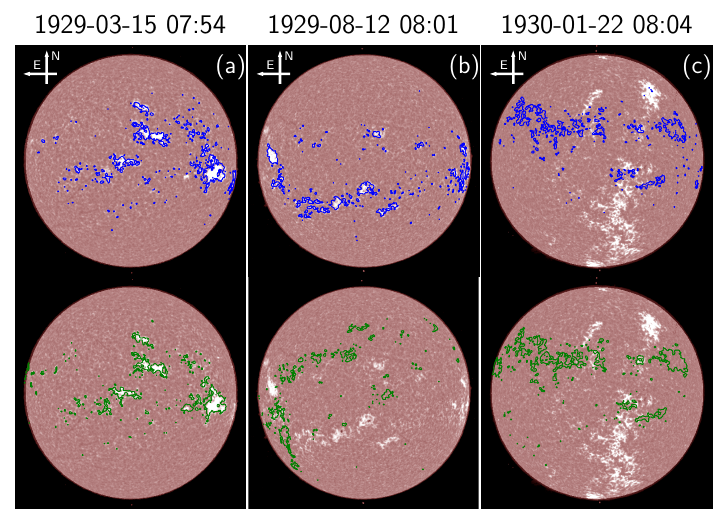}}
\caption{Examples of the procedure to identify mistakes in the observational date/time of the images.
The observations were made on 15 March 1929 at 07:54 \ac{ist}, 12 August 1929 at 08:01 \ac{ist}, and 22 January 1930 at 08:04 \ac{ist}. We show the calibrated images in the range of [-0.4, 0.4] in contrast values. In each case we overplot contours of plage regions from an earlier observation (Im$_{-1}$, shown in blue in the top row) and a later observation  (Im$_{+1}$, shown in green in the bottom row). With these images we show three different cases of overlap: (a) both contours overlapping, thus we marked \tobs\ to be correct, (b) only contour of Im$_{-1}$ is overlapping, again suggesting \tobs\ is correct (but \tobs\ of Im$_{+1}$ is most likely wrong), and (c) none of the contours are overlapping hence \tobs\ was flagged as incorrect. }
\label{fig4:diff_cases} 
\end{figure}

\begin{enumerate}
    \item First, we selected an observation (say Im$_0$) from the archive and using the \tobs\ in the file name we rotated the image according to the pole angle calculated using \autoref{eq2}.

    \item After that, we selected the immediately preceding, Im$_{-1}$) and immediately following, Im$_{+1}$, observations and rotate them accordingly with the same method. While selecting Im$_{-1}$ and Im$_{+1}$, we make sure that the difference in \tobs\ with Im$_0$ is between 0.5 to 3~days, to achieve as best comparison as possible.
    
    \item In the next step, we differentially rotated\footnote{Using IDL SolarSoftWare (SSW) routine {\it drot\_map.pro}.} Im$_{-1}$ and Im$_{+1}$ forward and back, respectively, to the time (\tobs) of Im$_0$ and over-plotted the contours of bright regions, using a fixed intensity threshold (see \autoref{fig4:diff_cases}a--\ref{fig4:diff_cases}c).
    
    \item In the last step, we manually checked the overlap of the plage contours from the various images to that of Im$_0$. If either of the contours shows a good overlap then we flag the \tobs\ (of Im$_0$) as correct (within the uncertainty of this exercise) otherwise we flag it as potentially incorrect.
\end{enumerate}

Three examples of this procedure are shown in Fig. \ref{fig4:diff_cases}.
In particular, the first two cases suggest that the observational date and time of Im$_0$ is correct since at least one of the contours from Im$_{-1}$ and Im$_{+1}$ exhibits a good overlap with the plage regions of Im$_0$. However, in \autoref{fig4:diff_cases}c, we see that none of the contours shows a good overlap. In general in this case besides the possibility that Im$_0$ has the wrong \tobs\ it is also possible that the Im$_0$ has correct \tobs\ but both the Im$_{-1}$ and Im$_{+1}$ have incorrect \tobs. Therefore to avoid such cases, we marked all three images as potentially having the wrong \tobs\ and we repeated the above mentioned process for all the observations marked in this step as having potentially incorrect \tobs. 
When repeating this, we made sure that, this time, we only choose Im$_{-1}$ and Im$_{+1}$, which have been previously flagged as having correct \tobs.

%
\begin{acks}
Kodaikanal Solar Observatory (KoSO) is a facility of the Indian Institute of Astrophysics, Bangalore, India. These data are now available for public use at \url{http://kso.iiap.res.in} through a service developed at IUCAA under the Data Driven Initiatives project funded by the National Knowledge Network. The authors also thank the observers at the KoSO for their meticulous work at maintaining this archive. BKJ expresses his gratitude to Max Planck Institute of Solar System and Research for the warm hospitality and financial support during his visit.
This project has received funding from the European Research Council (ERC) under the European Union's Horizon 2020 research and innovation programme (grant agreement No. 101097844 — project WINSUN).
T.C. thanks ISSI for supporting the International Team 474 “What Determines The Dynamo Effectivity Of Solar Active Regions?”. This research has made use of NASA's Astrophysics Data System (ADS; \url{https://ui.adsabs.harvard. edu/}) Bibliographic Services.

\end{acks}

\begin{authorcontribution}
BKJ devised the method to orient the observations and corrected the files for mistakes in observational dates/times, while also performing the bulk of the analysis presented in this study. TC processed the observations to produce the butterfly diagram as well as performed the cross-correlation technique to fix residual issues with the orientation.
The manuscript was mostly written by BKJ and TC, while all co-authors provided ideas, feedback about the work and proof read the manuscript.
\end{authorcontribution}

\begin{dataavailability}
The \ca\ data analyzed in this work is available through KoSO website at \url{http://kso.iiap.res.in}. The calibrated version of \ca\ data can be provided on request. The updated KoSO sunspot area series used is available through \url{https://dataverse.harvard.edu/dataset.xhtml?persistentId=doi:10.7910/DVN/JIST0V} \citep{kosoareav2}. The corrected \tobs\ and correct orientation of the \ca\ observation, fractional plage area (used for producing \autoref{fig4:butterfly}), the yearly latitudinal distribution (for \autoref{fig5:lat_ext}) along with other analyzed data are available at \url{https://github.com/bibhuraushan/KoSOCaK} and will be made available through Zenodo.
\end{dataavailability}
\begin{ethics}
\begin{conflict}
The authors declare no competing interests.
\end{conflict}
\end{ethics}

%
%
\bibliographystyle{spr-mp-sola}
\bibliography{references}  
%
%
%
%

\end{document}